\documentclass[submission]{eptcs}

\providecommand{\event}{MBT 2014} 
\usepackage{breakurl}             
\usepackage{graphicx}
\usepackage{tabularx}
\usepackage{amsmath}
\usepackage{algorithm2e}
\usepackage{enumitem}
\usepackage{fancyvrb}
\newtheorem{example}{Example}

\title{Verifying Web Applications: From Business Level Specifications to Automated Model-Based Testing}
\author{
Christian Colombo \qquad\qquad Mark Micallef \qquad\qquad Mark Scerri
\institute{PEST Research Lab\\
Department of Computer Science\\
University of Malta, Malta}
\email{\quad \{christian.colombo\,$|$\,mark.micallef\,$|$\,msce0013\}@um.edu.mt}
}
\def\titlerunning{Verifying Web Applications}
\def\authorrunning{C. Colombo, M. Micallef \& M. Scerri}

\begin{document}
\maketitle

\begin{abstract}
One of reasons preventing a wider uptake of model-based testing in the industry is the difficulty which is encountered by developers when trying to think in terms of properties rather than linear specifications.  A disparity has traditionally been perceived between the language spoken by customers who specify the system and the language required to construct models of that system.  The dynamic nature of the specifications for commercial systems further aggravates this problem in that models would need to be rechecked after every specification change.  In this paper, we propose an approach for converting specifications written in the commonly-used quasi-natural language Gherkin into models for use with a model-based testing tool.  We have instantiated this approach using QuickCheck and demonstrate its applicability via a case study on the \emph{eHealth} system, the national health portal for Maltese residents.
\end{abstract}

\section{Introduction}
\label{sec:introduction}

In typical commercial contexts, modern software engineering is characterised by a strong customer focus and, as a result, fluctuating requirements.  The industry has developed ways to adapt to this, most notably through Agile development processes \cite{COCKBURN:2001:AGILE} which have gained substantial popularity over the past decade. 
These processes start off by gathering requirements from the customer in the form of case scenarios and subsequently development is built around these scenarios culminating in their 
implementation and demonstration to the customer. This process,  
better known as behaviour-driven development \cite{CHELIMSKY:2010:RSPEC}, reorientates the development process so as to ground all work around business level customer specifications.  
Once the customer scenarios are implemented, the process tends to take on an iterative form whereby the customer is again asked to review the specification and possibly introduce further scenarios. 
To keep the specification language as simple as possible for non-technical customers, scenarios are typically specified in terms of a quasi-natural language called Gherkin \cite{GHERKIN}. In order to automate test execution, clauses from the Gherkin specification are associated with executable code using a technology such as Cucumber \cite{DEES:2013:CUCUMBER}.  This executable code effectively turns the specification into an automated acceptance test which is used to test that the system adheres to the customer's specifications. 

Whilst the automation achieved through the running of Gherkin specifications has significantly shortened the test execution process and made it repeatable, a significant limitation is that the scenarios are never executed in combination with one another: one would typically expect some classes of problems to arise  through the interactions between a sequence of scenarios rather than within the context of individual scenarios. 
In this context, we present an approach to combine related scenarios into models which can be consumed by model-based testing tools. In turn, such tools can be used to explore the model, generating plausible sequences of scenarios that provide more thorough testing of the system for minimal extra effort. 
We show how this technique has been instantiated on the QuickCheck testing tool and applied to Malta's national health portal web application.  This paper makes three main contributions:

$(i)$~A technique for automatically generating system models from existing test case scenarios written in the Gherkin language. These models can be consumed by model-based testing tools in order to generate test cases that extend normal testing and facilitate inter-scenario interaction.  
$(ii)$~A proof-of-concept tool which applies the technique presented here to generate models for the QuickCheck test case generation tool. 
$(iii)$~A case study on Malta's national health portal which demonstrates that the technique can be applied to a real-world system.


In order to control scope at this stage of our research, we constrain our work to the context of web applications.  Whilst this is a plausible constraint due to the increased proliferation of web applications, it does not preclude the technique from being modified and applied to other contexts.
The rest of the paper is organised as follows: Section \ref{sec:background} provides the reader with the background required to understand the work presented in this paper whilst Sections \ref{sec:approach} and \ref{sec:instantiation} explain the proposed technique and provides an overview of the instantiation respectively.  This is followed by Section \ref{sec:evaluation} whereby we discuss a case study designed to demonstrate the applicability of the technique.  Finally, Section \ref{sec:related_work} discusses related work from the literature and Section \ref{sec:conclusions} features our conclusions and opportunities for future work.


\section{Background}
\label{sec:background}

Gherkin is currently an industry standard language enabling non-technical customers to express requirements in terms of case scenarios. Ensuring that the developed system adheres to these specifications is crucial for customer acceptance of the resultant system. By associating executable code to statements within Gherkin, developers can use tools such as Cucumber to automatically execute test scenarios on the underlying system. However, to enable the automated manipulation of web applications through a browser, mimicking the user, one would need a tool such as Selenium which provides the necessary interface to access a web application programmatically. The rest of this section provides the necessary detail on these technologies so as to enable the reader to understand the work presented here.

\subsection{Gherkin}

Gherkin \cite{GHERKIN} is a business readable, domain-specific language which lets users describe what a software system should do without specifying how it should do it.  The language forms an integral part of \emph{Behaviour Driven Development} \cite{CHELIMSKY:2010:RSPEC}, a development process which aims to make specification a ubiquitous language throughout a system's life cycle.  Gherkin specifications take on a so-called \emph{Given-When-Then} format as illustrated in Example \ref{ex:featurespec} and the underlying concept is that a business user can specify what is expected from a system using a series of concrete scenarios.  \emph{Given} some precondition, \emph{When} the user does something specific, \emph{Then} some postcondition should materialise.  


\footnotesize
\begin{example}
\label{ex:featurespec}
\begin{Verbatim}[fontfamily=courier,fontshape=n,commandchars=\\\{\}]

 1.{\bf Feature:} Doctor's landing page
 2.{\bf Scenario:} Navigate to lab results page
 3. Given I am on the doctors landing page
 4. When I click on laboratory results
 5. Then I should go to the lab results page
 6.{\bf Scenario:} Navigate back to Doctor's landing page
 7. Given I am on the lab results page
 8. When I click on the myHealth logo
 9. Then I should go to the doctors landing page
\end{Verbatim}

\noindent In this example, the client has specified two requirement scenarios as part of a feature on an eHealth system: one dealing with the process of a doctor viewing the laboratory results and another one of a doctor navigating back to the landing page.  
\end{example}

\normalsize

Developers then decompose each scenario into a number of development tasks and the work proceeds with all activity being a direct result of a particular customer-specified scenario.  Furthermore, although the scenarios are written in a subset of natural language, technologies exist (e.g.\ Cucumber \cite{DEES:2013:CUCUMBER} and Specflow \cite{SPECFLOW:ONLINE}) which enable developers to implement automated user acceptance tests for every step in a scenario.  That is to say that 
$(i)$ the user specification becomes executable as an automated test by the end of a development iteration; and 
$(ii)$ any future scenarios which include a predefined step will automatically reuse test automation code, thus in effect building up a domain specific language for automated testing.

\subsection{Cucumber}



Cucumber \cite{DEES:2013:CUCUMBER} is a technology that enables developers to provide glue code which turns natural language specifications into executable automated tests.  Although originally written in Ruby, the tool has been ported to a number of languages.  It reads plain language text files called \emph{feature files} (as in Example~\ref{ex:featurespec}), extracts scenarios and executes each scenario against the system under test.  This is achieved by means of so-called \emph{step definitions}, a series of methods which are executed when a step in a feature file matches a particular pattern. For example for the statement \verb+When I click on the myHealth logo+, the developer associates code which actually simulates a user clicking on the logo. 
Once all Gherkin statements have associated code which can be executed by Cucumber, the latter will iterate through each step, find a matching step definition and execute the method associated with it.  Step definitions can implement any feature supported by the underlying language.  In our case, we are interested in manipulating web browsers for automated web application testing; something which can be achieved by Selenium.


%
%

\subsection{Selenium}

Selenium \cite{SELENIUM} is a web browser automation framework which has become a de-facto industry standard when it comes to web test automation.  It has been implemented in a number of popular programming languages and provides API-level access to various browsers.  In essence, developers can write programs which launch a browser,  simulate user behaviour within the browser and subsequently query the state of website elements to verify expected behaviour as illustrated in Example \ref{ex:selenium}.  

\footnotesize
\begin{example}
\label{ex:selenium}
\begin{Verbatim}[fontfamily=courier,fontshape=n,commandchars=\\\{\}]

public void testGoogle() \{
   {\bf // Launch browser and visit www.google.com}
    WebDriver browser = new FireFoxDriver();
    browser.open("http://www.google.com"); 
   {\bf // Search for ``Hello World"}
    browser.findElement(By.id("q")).sendKeys("Hello World");
    browser.findElement(By.id("btnG")).click(); 
   {\bf // Check that the search engine behaved appropriately}
    assertEqual("Hello World - Google Search", browser.getTitle());
\}
\end{Verbatim}
\noindent The above code snippet features source code written in Java which carries out a test on the Google search engine using the Firefox browser.  The \verb+WebDriver+ object in this example is provided by the Selenium framework.
\end{example}
\normalsize

Combining Cucumber with Selenium can result in truly powerful automated test suites which provide repeatable and consistent regression testing capabilities.  We argue that this combination can be leveraged even further by bringing model-based testing into the mix at minimal cost by constructing models from existing Gherkin specifications.

\section{Proposed Technique}
\label{sec:approach}



Starting with a number of Gherkin scenarios, the proposed technique combines characteristics inherent in the \emph{Given-When-Then} notation with those of web applications 
in order to make educated inferences and consequently construct a model of a specified system.  In order to construct a model representation of web applications, the specification language should allow us to:

\begin{enumerate}
\item Define the states of the system
\item Specify the starting state
\item Specify the transitions available at each state, including any preconditions, the action which causes the system to change state, and any postconditions
\end{enumerate}

We argue that the \emph{Given-When-Then} structure of Gherkin lends itself to providing this information since a single scenario in \emph{Given-When-Then} form can be interpreted to represent a single transition between two states.  That is to say that the \emph{Given} step indicates an origin state (and possibly preconditions), the \emph{When} state specifies actions for a transition to occur, 
and finally the \emph{Then} step indicates the target state (possibly with postconditions).  
However, in order to facilitate the automated interpretation of scenarios in this manner, we propose that users of the language adopt three conventions taking into account  the navigation-driven nature of web applications.  
To illustrate the conventions, we will be using Example \ref{ex:featurespec2} as a running example.


\footnotesize
\begin{example}
\label{ex:featurespec2}
\begin{Verbatim}[fontfamily=courier,fontshape=n,commandchars=\\\{\}]

 1.{\bf Feature:} Patient management features
 2.{\bf Scenario:} Navigate to lab results page
 3. Given I start on the "doctors landing page"
 4. And I have pending lab results
 5. When I click on laboratory results
 6. Then I should go to the "lab results page"
 7. And I should see my pending lab results
 8.{\bf Scenario:} Navigate back to Doctor's landing page
 9. Given I am on the "lab results page"
10. When I click on the myHealth logo
11. Then I should go to the "doctors landing page"
\end{Verbatim}

\noindent The above Gherkin code consists of two scenarios which are authored according to the conventions proposed here. More details on the pattern of these scenarios are given below when describing the conventions. 

\end{example}
\normalsize

\noindent Corresponding to the three points outlined about, we propose the conventions below:

\begin{description}
\item [Convention 1: Make states explicit.]  In order to identify states within a test scenario, we constrain the language such that \emph{Given} steps take the form of \textbf{\emph{Given I am on the ``\textless web page \textgreater''}}.  Similarly, \emph{Then} steps take the form of \textbf{\emph{Then I should be on the ``\textless web page \textgreater''}}.  Recall that we consider web pages within an application to represent states within the application's model.  This convention, whereby \emph{Given} and \emph{Then} steps take a specific form which includes state names being surrounded by quotes, enables us to clearly identify states within the model.  In Example \ref{ex:featurespec2}, the states ``doctors landing page'' and ``lab results page'' can be extracted from lines 3, 6, 9 and 11.  Upon closer inspection, one realises that line 3 takes on a slightly different form.  This is outlined in the next convention.

\item [Convention 2: Identify start states.]  In order to identify start states in a model, we propose that the convention on \emph{Given} steps consist of two variations.  The first is the \textbf{\emph{Given I am on the ``\textless web page \textgreater''}} outlined in \emph{Convention 1} whereas the second would replace ``I am'' with ``I start'' resulting in \textbf{\emph{Given I start on the ``\textless web page \textgreater''}}.  This minor variation will identify the web page in question as a starting state in the system's model.

\item [Convention 3: Identify actions, preconditions and postconditions.]  Finally, we identify the \emph{When} construct as the action which takes the user from one page to another while the \emph{And} steps which form part of \emph{Given} or \emph{Then} parts of scenarios will be leveraged to identify preconditions and postconditions respectively. 
That is to say that this additional information can be used to infer conditions which should be satisfied before a transition can fire, as well as conditions which should be satisfied after a transition has fired. 
Consider the first scenario in Example \ref{ex:featurespec2}. 
The action taking the user from the ``doctors landing page'' to the ``lab results page'' is identified by the \emph{When} clause:  ``clicking on laboratory results''. 
Furthermore, in this case, the \emph{Given} portion of the scenario has two lines.  The first identifies the current state as the ``doctors landing page'' whilst the second sets up a precondition on the subsequent transaction.  In order for ``clicking on laboratory results'' to be a valid action, first the condition that the doctor has pending lab results must be satisfied.  Similarly the \emph{Then} portion of the scenario identifies the ``lab results page'' as the target state of the scenario whilst line 7 states that the doctor should also see pending lab results within that page once the transition is complete.
\end{description}

\noindent If these conventions are adhered to, test scenarios can be interpreted as follows:

\begin{description}
\item [Given:] The \emph{Given} step is usually used to specify preconditions in a test scenario but in the context of model-based testing, it can be used to infer the current state of the system. In accordance to convention 2, a special adaptation of this step can also be used to specify the starting state of a model.  Furthermore, any \emph{And} steps immediately following the \emph{Given} step will be treated as preconditions.
\item [When:] In normal use, the \emph{When} step specifies actions for exercising the system under test in a scenario and retains this role in our technique such that the action specifies a transition from the current state to a target state. 
\item [Then:] The \emph{Then} step is usually utilised to specify a postcondition in a test scenario but in our technique we adapt this to specify the target state.  Furthermore, any \emph{And} steps immediately following the \emph{Then} step will be treated as postconditions.
\end{description}

\begin{figure}
\centering{{\includegraphics[width=0.8\textwidth]{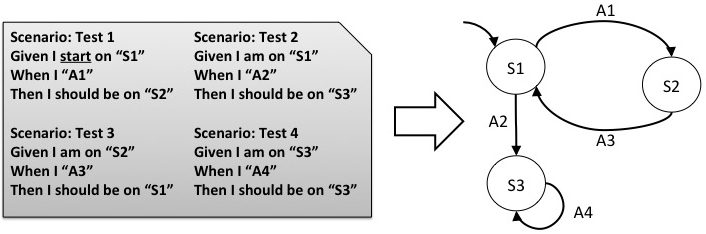}}}
\caption{A depiction of how models can be constructed from multiple test scenarios.}
\label{fig:generic_example}
\end{figure}

\begin{figure}
\centering{{\includegraphics[width=0.62\textwidth]{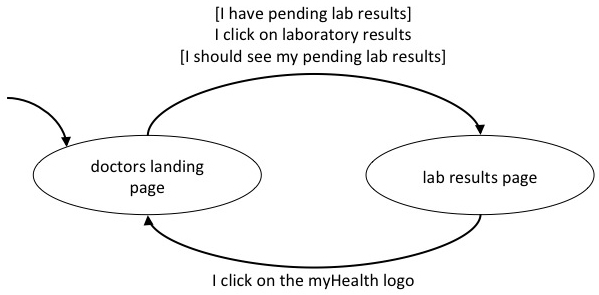}}}
\caption{A model constructed by applying Algorithm \ref{algorithm} to Example \ref{ex:featurespec2}.}
\label{fig:model-ex2}
\end{figure}

By processing multiple test scenarios and joining them at points where states from different scenarios have the same name, a model can now be constructed. 
This is demonstrated in Figure \ref{fig:generic_example} whereby four generic test scenarios are combined to construct a single model. 
More formally, given a list of test scenarios written in Gherkin such that they adhere to the conventions discussed above, a model can be constructed using Algorithm \ref{algorithm}. In short the algorithm processes the scenarios, identifies unique states and inserts the transitions including actions, preconditions and postconditions. Figure \ref{fig:model-ex2} shows a model constructed by applying this algorithm to Example \ref{ex:featurespec2}.


In the next section, the approach described above is instantiated in the context of a particular model-based testing tool. 

\begin{algorithm}
 \SetAlgoLined
 \KwData{scenarios : List of Test Scenarios}
 \KwResult{model : A finite state machine modelling the system under test.}
model = new Model()\;
 \While{scenarios.hasMoreScenarios()}{
  scenario = scenarios.nextScenario()\;
  os = scenario.extractOriginState() \tcp*{According to Convention 1}
  ts = scenario.extractTargetState() \tcp*{According to Convention 1}
   \eIf{(os != null $\wedge$ ts != null)} {
     model.addStateIfNew(os)  \;
     model.checkIfStarting(os)              \tcp*{May be marked as start state if Convention 2}
     model.addStateIfNew(ts) \;
     preconditions = scenario.extractPreconditions()   \tcp*{According to Convention 3}
     postconditions = scenario.extractPostconditions() \tcp*{According to Convention 3}
     actions = scenario.extractActions() \tcp*{All actions from the When steps}
     \eIf{(actions != null)} {
       transition = new Transition(preconditions, actions, postconditions)\;
       model.addTransition(os,ts,transition)\;
     } {
       error(Unable to extract transition from test scenario)\;
     }
  }{
   error(Unable to extract states from test scenario)\;
  }
  return model;
 }
 \caption{High level algorithm for generating models from business specifications.}
 \label{algorithm}
\end{algorithm}

\section{Instantiation}
\label{sec:instantiation}

As a prerequisite to evaluating the approach, we developed an instantiation: namely stringing together three parts: $(i)$ the algorithm translating the Gherkin scenarios into a model for a model-based testing tool, $(ii)$ the model-based testing tool which generates the test cases and detects test failure or success, and $(iii)$ Selenium which enables the testing tool to interact directly with the web application. 

\begin{figure}
\centering{{\includegraphics[width=0.55\textwidth]{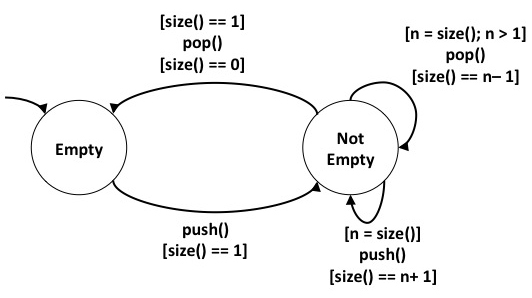}}}
\caption{An example of a model specified in QuickCheck.}
\label{fig:model-ex1}
\end{figure}

While other model-based testing tools such as ModelJUnit\footnote{\texttt{http://www.cs.waikato.ac.nz/\~{}marku/mbt/modeljunit/}} are perfectly valid alternatives, as a model-based testing tool, we selected QuickCheck \cite{CLAESSEN:2011:QUICKCHECK} for the fact that it provides test case \emph{shrinking} functionality whereby upon detecting a failure, it searches for smaller test cases which also generate the fault to aid debugging.
Although initially developed as a tool for the Haskell programming language, it has been implemented on a number of platforms and for this work, we utilise the Erlang implementation \cite{ARTS:2006:QUVIC}.  
QuickCheck is primarily a random test data generation tool but it also supports model-based testing. Given a model and a set of pre and postconditions, the tool is able to automatically generate test cases which are used to verify that the conditions hold in the context of different sequences of actions.  Figure~\ref{fig:model-ex1} illustrates an example QuickCheck model which represents valid sequences of operations involving a stack (top elements in square brackets represent preconditions while the bottom elements represent postconditions). 
Elements can be pushed or popped from the stack: Popping an element from the stack is not allowed when the stack is empty, otherwise the size should be decremented by one. On the other hand, pushing an element should result in the size of the stack to increase by one. 

A QuickCheck test case execution starts from the initial state and selects a transition whose precondition is enabled. The action upon the transition is actuated and the postcondition is checked. If a postcondition check fails, then this signifies a failed test.

\begin{figure}
\centering{{\includegraphics[width=0.75\textwidth]{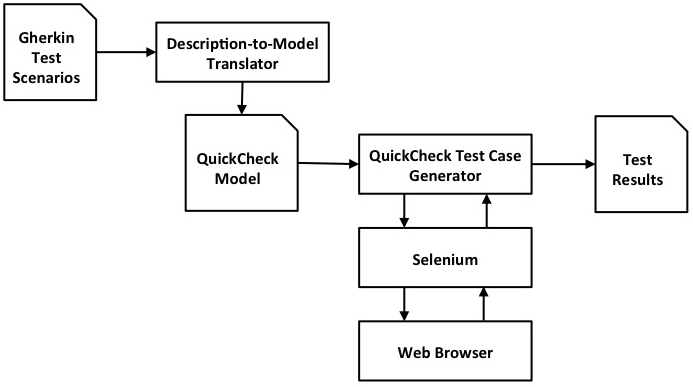}}}
\caption{An overview of the system architecture}
\label{fig:architecture}
\end{figure}

Once QuickCheck has been chosen to instantiate our approach, a tool was developed as illustrated in the architecture diagram in Figure~\ref{fig:architecture}.  The system takes a set of test scenarios as input and proceeds to parse them and convert them into a QuickCheck model.  This model is then fed to QuickCheck for test case generation and execution.  During the test case execution process, Selenium is utilised for interacting with a web browser and exercising the web application.  Finally, QuickCheck outputs a report containing results of the test run.

Having implemented an instantiation of the technique, we could explore its applicability through an industry case study.  This is the subject of the following section.

\section{Case Study}
\label{sec:evaluation}
Our initial evaluation of the technique presented here is a case study.  The objectives of this case study were threefold: 

\begin{enumerate}
\item Demonstrate the feasibility of the technique on a non-trivial system
\item Compare the technique to its manual alternative
\item Note any relevant observations to help form future research directions
\end{enumerate}

The case study was carried out on myHealth\footnote{\texttt{http://www.myhealth.gov.mt}}, a web-based solution that has recently been developed for the government of Malta's Ministry of Health to provide an improved and more efficient service to patients in the Maltese health care infrastructure.  It consists of a web portal allowing access to myHealth data for registered citizens.  A back-end portal is provided so as to allow monitoring and management to administrators.  The case study was scoped to the front-end portal only, mainly because the back-end portal was purchased off the shelf and no test scenarios were available for it.  The system is able to distinguish between doctors and citizens, providing different functionality for the two roles with Table \ref{table:functionality} providing a list of high level functionality available to both types of users.  When citizens gain access to their myHealth Record account, they are able to search for doctors listed in the Medical Council Register, request a specific doctor to be their \emph{trusted doctor} and go on to view and manage all their personal health data.  

\begin{figure}
\centering{{\includegraphics[width=0.75\textwidth]{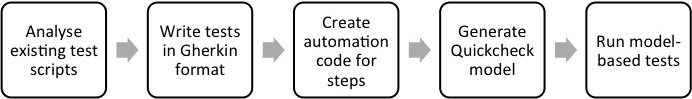}}}
\caption{The process followed when carrying out the case study.}
\label{fig:casestudy_process}
\end{figure}

The case study was carried out as depicted in Figure \ref{fig:casestudy_process}.  Since the system's test scripts were not documented in Gherkin format, they needed to be analysed and written in the required form.  Once this was done, test automation code was created such that all the scenarios became executable automated tests.  This enabled us to feed the test scenarios into our tool, create QuickCheck models and subsequently generate and execute tests based on the models. 
These steps are elaborated in the following subsections. 

\begin{table}
  \begin{center}
  \begin{tabular}{|p{0.45 \textwidth}|p {0.45 \textwidth}|}
   \hline
   \textbf{Actions available to doctors} & \textbf{Actions available to citizens} \\
   \hline
   \begin{itemize}[topsep=-0.5cm,leftmargin=0.4cm]
   \item Login as a doctor
   \item Go to \emph{appointments} page
     \begin{itemize}
       \item Search for a patient's appointments
       \item View a patient's appointments
     \end{itemize}
   \item Go to \emph{case summaries} page
   \begin{itemize}
       \item Search for a patient's case summary
       \item View a patient's case summary
       \item Print patient's case summary
     \end{itemize}
   \item Go to \emph{laboratory results} page
      \begin{itemize}
       \item Search for a patient's lab result
       \item View a patient's lab result
       \item Release result to patient
       \item Mark lab results as read or unread
       \item Print patient's lab results
     \end{itemize}
   \item Go to \emph{medical image reports} page
     \begin{itemize}
       \item Search for a patient's report
       \item View a patient's report
       \item Release report to patient
       \item Mark medical image reports as read or unread
       \item Print patient's medical image report
     \end{itemize}
    \item Go to \emph{POYC entitlement} page
     \begin{itemize}
       \item Search for a patient
       \item View a patient's entitlement
     \end{itemize}
     \item Go to \emph{search patients' data} page
      \begin{itemize}
       \item Search for a patient
       \item View a patient's records
     \end{itemize}
     \item{Go to \emph{manage patients} page}
      \begin{itemize}
       \item Search for a patient
       \item Remove a patient
     \end{itemize}
     \item Go to \emph{manage notifications} page
      \begin{itemize}
       \item Subscribe to email notifications
      \end{itemize}
   \end{itemize}
   
   & 
    \begin{itemize}[topsep=-0.5cm,leftmargin=0.4cm]
    \item Go to \emph{appointments} page
     \begin{itemize}
       \item View appointments
     \end{itemize}
     \item Go to \emph{case summaries} page
      \begin{itemize}
       \item Search for case summary
       \item View a case summary
       \item Print case summary
     \end{itemize}
      \item Go to \emph{laboratory results} page
      \begin{itemize}
       \item Search for a lab result
       \item View a lab result
       \item Mark lab results as read or unread
       \item Print lab results
     \end{itemize}
      \item Go to \emph{medical image reports} page
      \begin{itemize}
       \item Search for a report
       \item View a report
       \item Mark medical image report as read or unread
       \item Print medical image reports
     \end{itemize}
      \item Go to \emph{POYC entitlement} page
     \begin{itemize}
       \item View a POYC entitlement
       \item Print entitelment
     \end{itemize}
     \item Go to \emph{search for doctors} page
     \begin{itemize}
       \item View all registered doctors
       \item Search for a doctor
       \item Send request to a doctor
     \end{itemize}
     \item Go to \emph{manage doctors} page
       \begin{itemize}
       \item Remove a doctor
      \end{itemize}
   \item Go to \emph{manage notifications} page
         \begin{itemize}
       \item Subscribe to email notifications
         \item Subscribe to mobile notifications
      \end{itemize}
    \end{itemize}
   \\
   \hline
  \end{tabular}
  \caption{Functionality available to the doctors and citizens in the myHealth system.}
  \label{table:functionality}
  \end{center}
\end{table}

 \subsection{Constructing a model for the case study}

When we were granted access, the myHealth Record system had already been through testing and been launched to the public.  In this section we briefly outline the testing process used by the company that developed myHealth.  The system went through two phases of testing: the first phase was a unit testing and integration phase which was mainly carried out by developers in the company.  The second phase consisted of user acceptance testing and was carried out in conjunction with the customer in order to verify that the required functions had been implemented and were working correctly.  The nature of the work presented in this paper is mainly concerned with system-level or user acceptance testing, so the second phase is of interest to us.  This phase was driven by pre-documented test cases which outlined a series of procedures and expected outcomes (see Table \ref{table:testcase} for a sample test case).  

\begin{table}
  \begin{center}
  \begin{tabular}{|l|l|}
  \hline
  \textbf{Test Case Number:} & 4 \\
  \hline
  \textbf{Scope:} & Releasing the full Updated Report through the Patient Data menu \\
  \hline
    \multicolumn{2}{|l|}{\textbf{Procedure and Expected Outcome:}} \\
    \hline
    \multicolumn{2}{|p{0.9 \textwidth}|}{
    Doctor - Releasing the report
    \begin{enumerate}
      \item Login to the system as Doctor - e-ID 0240860M
      \item In the Doctor Landing Page, open the \emph{Patient Data} accordion
      \item Navigate to the \emph{Medical Image Reports} section in the accordion
      \item Locate the previous report and Click on \emph{View Reports}
      \item Click \emph{Release All}
      \item Logout
    \end{enumerate}

    Citizen - Testing that the updated report has been released
    
    \begin{enumerate}
    \item Login to the system as Citizen - e-ID 0240851M
    \item In the Citizen Landing Page, open the \emph{myMedical Image Reports accordion}. Note that the \emph{Updated box} in the accordion shows 1
    \item Locate a report and note that an \emph{Updated Reports} icon is displayed
    \item Click on \emph{View Reports} and note that some of the reports were updated.  Previous reports should be marked as \emph{Superseded} and greyed out while the updated report should be marked as \emph{Final (Updated)}. The report should contain the updates.
    \item Logout
    \end{enumerate}

    }\\
    \hline
    \textbf{Date Run:} & 01/03/2012 \\
    \hline
    \textbf{Run By:} & \textless Names of testers present\textgreater \\
    \hline
    \textbf{Overall Test Outcome:} & Pass \\
    \hline
     \multicolumn{2}{|l|}{\textbf{Additional Commands and/or Screenshots:}} \\
    \hline
    \multicolumn{2}{|l|}{ } \\
    \hline
  \end{tabular}
  \caption{A sample test case from the user acceptance testing phase}
  \label{table:testcase}
  \end{center}
\end{table}

\begin{figure}
\centering{{\includegraphics[width=0.85\textwidth]{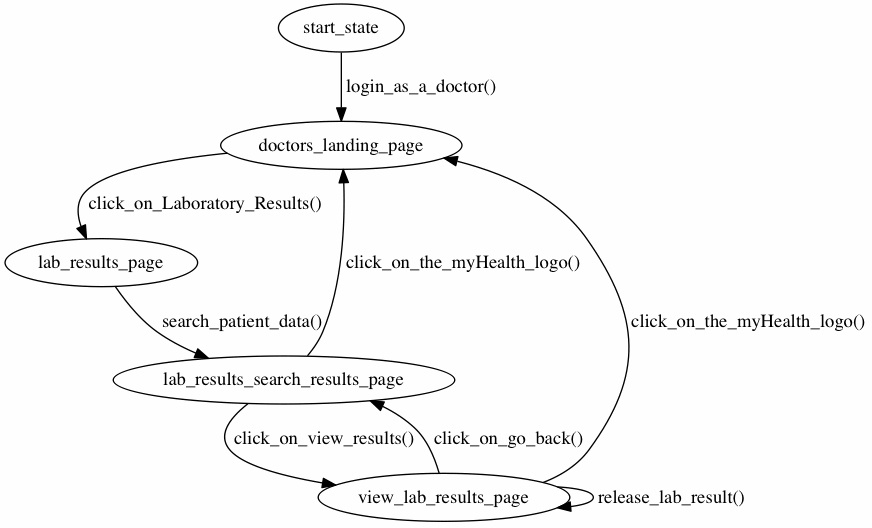}}}
\caption{Model generated for myHealth's \emph{Doctor's Lab Results} functionality}
\label{fig:casestudy_model}
\end{figure}

For the purposes of this case study, the test cases from the second phase of testing were secured and used as the basis to construct specifications in the Gherkin language.  Each test case was manually analysed and systematically converted to its equivalent \emph{Given-When-Then} format.  Furthermore, step definitions were implemented such that each scenario became an executable automated test.  Related test scenarios were then grouped together and models were generated accordingly using an implementation of Algorithm \ref{algorithm}.  Figure \ref{fig:casestudy_model} shows one of the models generated during the exercise.

\subsection{Results and Observations}

With regards to exploring the feasibility of such an approach on a non-trivial system, the case study demonstrated that this is indeed the case.  We were able to generate models directly from business level specifications and subsequently utilise those models to generate and execute test cases.  The process took approximately seven days of work to complete.  This was similar to the length of time taken by the company to carry out manual user acceptance testing of the product.  However, if one looks closer at how the time was invested (see Table \ref{table:timings}), in the case of manual testing, the company estimates that four days were spent designing test cases whilst three days were spent executing them.  
In the case of automated model-based testing, the test design aspect was practically free (due to the use of QuickCheck) whilst the actual execution only took one day.  Whilst on the face of it, it seems that both approaches are equally time consuming, it is worth noting that if the company had adopted the use of Cucumber in its normal operations (as is the case with an increasing number of companies), the time require for automated model-based testing would have been substantially reduced.  Furthermore, automated model-based testing is efficiently repeatable and will deliver consistent results when compared to that delivered manually by human testers.  

Two main observations were made throughout this process: Firstly, automated testing will not be able to replicate all tests which are carried out manually by a human tester.  In fact, we were able to replicate 74\% of manual tests due to test cases which either required human judgement (e.g.\ checking printouts) or were too complex to feasibly automate (e.g.\ temporal tests).  The second observation was related to the shrinking feature of QuickCheck: while this particular feature is fast in console-style applications, we observed that shrinking took an inordinately long amount of time when testing web applications.  This is mainly due to the expense required to repeatedly launch a browser and navigate through a series of web pages during the shrinking process.

\begin{table}
  \begin{center}
  \begin{tabular}{l|c|c|}
  \cline{2-3}
  & \textbf{Manual} & \textbf{Automated} \\
  & \textbf{Testing} & \textbf{Model-based Testing} \\
  \hline
  \multicolumn{1}{ |l| }{Generation of test cases} & 4 days & - \\
  \hline
   \multicolumn{1}{ |l| }{Creating and automation Gherkin Scenarios} & - & 6 days \\
   \hline
    \multicolumn{1}{ |l| }{Execution of test cases} & 3 days & 1 day \\
    \hline
     \multicolumn{1}{ |l| }{\textbf{Total time}} & 7 days & 7 days \\
     \hline
  \end{tabular}
    \caption{Comparison of the time required for manual testing and automated model-based testing.}
  \label{table:timings}
  \end{center}
\end{table}

It is worth noting that no new bugs were uncovered as a result of applying the technique.  We think that this is due to the fact that by the time we ran the case study, the system had been thoroughly tested, deployed and in active use for quite some time.  

\subsection{Threats to Validity}

The evaluation presented here is mostly concerned with demonstrating that the automated generation of models from natural language specifications and the subsequent generation and execution of tests is prima facie feasible.  Consequently, the case study is limited in terms of depth and is subject to a number of external validity threats.  We split these threats into two categories: $(i)$~threats related to new challenges introduced by this technique and $(ii)$~threats related to test engineering challenges which are manifested in most automation attempts.  We argue that the first category of threats are highly plausible and should be investigated further whilst the second category of threats is diminished due to the fact that we are reusing existing automation code which should have already dealt with them.

The main threat introduced by the current approach is concerned with the requirement that conventions be followed when constructing natural language specifications.  That is to say that preconditions, actions and post conditions are assumed to reside in specific elements within each scenario.  Whilst the conventions are arguably reasonable, they do introduce some restrictions on what is meant to be a flexible language.  This constitutes a threat to validity because unlike this case study whereby scenarios were created for the purpose of the evaluation, real-life scenarios will be written by different people who may or may not have followed the same conventions.  

With regards to test engineering challenges, two representative issues come to mind.  Firstly, one can also argue that the data used in this case study was not sufficiently complex and in reality, data dependencies may be highly complex.  Whilst the concern is justified, we argue that the threat to validity here is limited, mostly due to the fact that the technique assumes that an automated testing infrastructure already exists.  Such infrastructures usually contain sophisticated test data factories which support automated testing.  These factories are typically called by `\emph{Given}' steps in order to create/fetch any required data as well as by `\emph{Then}' steps to tear down and residual test data that is no longer required.   Similar arguments can be made for other classes of test engineering challenges.   For example, some test could be sensitive to timing issues.  However, since we are essentially reusing existing test automation code which is exposed to the same challenges, it can be safely assumed that test engineers would have solved these challenges within that code.  If not, then the original set of automated test would not be functional in the first place.  Whilst this should reduce validity concerns related to data dependencies, we do intend to observe this in more elaborate case studies as discussed in Section \ref{sec:conclusions}.

\section{Related Work}
\label{sec:related_work}
The application of model based testing to web applications is not new and has been proposed on numerous occasions.  Andrews et al.\ \cite{ANDREWS:2005:MBTWEB} proposed the use of finite state machines as an ideal model for this domain.  Marchetto et al.\ \cite{MARCHETTO:2008:STATE} extended this approach to cater for more dynamic Ajax-based applications. 
Similarly, Wu and Offutt \cite{wu2002modeling} propose a technique to model both static and dynamic aspects of web applications based on identifying elements of dynamically created web pages and combining them using sequence, selection, aggregation and regular expressions to create composite web pages. 
Ernits et al.\ \cite{ERNITS:2009:MBT} demonstrate how model based testing can be applied to web applications using the NUnit tool.  NUnit allows developers to create models in C\# and subsequently uses those models to generate test cases.  Whilst the authors had success with the technique, they commented that a significant amount of time was spent learning the modelling formalism and building the test harness \cite{ERNITS:2009:MBT}. This contrasts to our approach which bypasses the need of manually programming the models as long as the business-level specifications adhere to the conventions highlighted in this paper.

With regards to automatic model generation, the literature mostly reveals work utilising reverse engineering approaches.  Hajiabadi and Kahani \cite{hajiabadi2011automated}  propose a technique for first modelling the structural aspects of a web application by observing client requests and server responses, and subsequently applying ontologies to generate test data.  The authors claim that ontologies are required in order to make test case selection more effective. Our approach differs in that ours focus on the generation of sequences of actions rather than the data used. (This is left as future work in our case.) 
Ricca and Tonella \cite{ricca2001analysis} propose a UML representation of web applications and present a tool called \emph{ReWeb} which reverse engineers web applications into this notation.  They also present a tool named \emph{TestWeb} which processes UML representations of web sites to automatically generate and execute test cases. While this approach is similar to ours in the use of a high-level specification notation, it differs in that it uses the implementation of the system itself to extract the model while we use the high-level business specification to this end. 

The closest work to ours is that carried out by J{\"a}{\"a}skel{\"a}inen et al.\ \cite{JKK+09synthesizing} which similarly synthesises and merges test models from simple linear test cases. The main difference 
is that our propositions are not state-based but rather transition-based, resulting in simpler states. The implication is that our merging is fully automatable (with the possibility of having duplicate transitions) while in their case the merging has to be manually checked. 


\section{Conclusions and Future Work}
\label{sec:conclusions}

In this paper we proposed an approach for converting business level specifications written using \emph{Given-When-Then} natural language notation into models that can subsequently be used to generate and execute test cases.  Whilst the literature contains multiple proposals for model-based testing of web applications, most techniques require the involvement of highly technical individuals who were capable of constructing models of their systems. Given the dynamic evolutionary nature of modern systems, maintaining models in the long run may not be desirable or even feasible.   Our technique can work off natural language specifications which adhere to three simple conventions as discussed in Section \ref{sec:approach} and if applied in a company where automated testing is already the norm, model creation and subsequent test case generation can be achieved at minimal cost.

With regards to future work, there are a number of paths which we would like to pursue:
 
 \begin{itemize}
 
 \item
  The first involves converting the technology to make use of different model-based testing tools on various platforms.  This is desirable because the most expensive part of the approach is the automation of test scenarios in the language of the model-based testing tool, which in our case was Erlang.  If models can be generated in the same language as that being used by developers in a particular company, then test automation code can simply be reused by models out of the box.  
  
  \item
  The second path of research involves carrying out larger case studies both in terms of system size, as well as in terms of project duration.  It would be ideal to find industry partners who have ongoing projects using Cucumber as a test automation tool and would like to utilise our technique on a long term basis.  This would help us make more detailed observations and modify the technique accordingly.
  
  \item
  While our approach is a perfect fit for web applications due to the ease with which scenarios can be connected over common web pages, we are confident the approach can also be applied to other domains such as desktop GUI-based applications where the connecting element might be a tab or a window which is enabled at a particular point in time. 
  
\item  
  On a more theoretic level, we wish to explore the possibility of modelling variations of data elements rather than limiting ourselves to sequences of actions. For example in the case study we could model different laboratory results, doctor attributes, patient conditions, etc., automatically to generate more varied test cases.
  
  \item
  Finally, it is rarely the case that scenarios run in a sequential fashion in a real life environment. Thus, it would also be interesting to explore models which allow modelling several users concurrently executing different scenarios.
  
  \end{itemize}

\bibliographystyle{eptcs}
\bibliography{references}

\end{document}